\documentclass[prd,superscriptaddress,nofootinbib,amsfonts,notitlepage,longbibliography]{revtex4-2}
\usepackage[utf8]{inputenc}
\usepackage[T1]{fontenc}
\usepackage{hyperref}
\usepackage{graphicx,bm,amsmath,color,scalerel}
\usepackage{bbold}
\usepackage{wasysym}
\usepackage{verbatim}
\usepackage{amssymb}
\usepackage{epstopdf}

\usepackage{animate}
\usepackage{xmpmulti}
\usepackage{centernot}
\usepackage{multirow}
\usepackage{tikz}
\usepackage{comment}
\usepackage[normalem]{ulem} 
\makeatletter

\newcommand{\be}{\begin{equation}}
\newcommand{\ee}{\end{equation}}
\newcommand{\beq}{\begin{eqnarray}}
\newcommand{\eeq}{\end{eqnarray}}
\newcommand{\ba}{\begin{align}}
\newcommand{\ea}{\end{align}}

\newcommand{\epsth}{\varepsilon^{\text{thr}}}

\begin{document}

\title{Modification of the mean free path of very high energy photons \\ due to a relativistic deformed kinematics}
\author{J.M. Carmona}
\email{jcarmona@unizar.es}
\affiliation{Departamento de F\'{\i}sica Te\'orica,
Universidad de Zaragoza, Zaragoza 50009, Spain}
\affiliation{Centro de Astropartículas y Física de Altas Energías (CAPA),
Universidad de Zaragoza, Zaragoza 50009, Spain}

\author{J.L. Cort\'es}
\email{cortes@unizar.es}
\affiliation{Departamento de F\'{\i}sica Te\'orica,
Universidad de Zaragoza, Zaragoza 50009, Spain}
\affiliation{Centro de Astropartículas y Física de Altas Energías (CAPA),
Universidad de Zaragoza, Zaragoza 50009, Spain}

\author{J.J. Relancio}
\email{relancio@unizar.es}
\affiliation{Departamento de F\'{\i}sica Te\'orica,
Universidad de Zaragoza, Zaragoza 50009, Spain}
\affiliation{Centro de Astropartículas y Física de Altas Energías (CAPA), Universidad de Zaragoza, Zaragoza 50009, Spain}
\affiliation{Dipartimento di Fisica ``Ettore Pancini'', Università di Napoli Federico II, 80138 Napoli, Italy}
\affiliation{INFN, Sezione di Napoli, 80126 Napoli, Italy}
\affiliation{Departamento de Física, Universidad de Burgos, 09001 Burgos, Spain}

\author{M.A. Reyes}
\email{mkreyes@unizar.es}
\affiliation{Departamento de F\'{\i}sica Te\'orica,
Universidad de Zaragoza, Zaragoza 50009, Spain}
\affiliation{Centro de Astropartículas y Física de Altas Energías (CAPA),
Universidad de Zaragoza, Zaragoza 50009, Spain}

\author{A. Vincueria}
\email{angelvincu@gmail.com}
\affiliation{Departamento de F\'{\i}sica Te\'orica,
Universidad de Zaragoza, Zaragoza 50009, Spain}

\begin{abstract}
Ultra-high-energy physics is about to enter a new era thanks to the impressive results of experiments such as the \emph{Large High Altitude Air Shower Observatory} (LHAASO), detecting photons of up to $1.4\times 10^{15}$\,eV (PeV scale). These new results could be used to test deviations with respect to special relativity.
While this has been already explored within the approach of Lorentz Invariance Violation (LIV) theories,  in this work we 
consider, for the first time, modifications due to a relativistic deformed kinematics (which appear in Doubly Special Relativity, or DSR, theories). In particular, we study the mean free path of very high-energy photons due to electron-positron pair creation when interacting with low-energy photons of the cosmic microwave background. Depending on the energy scale of the relativistic deformed kinematics, present (or near future) experiments  can be sensitive enough to be able to identify deviations from special relativity. 
\end{abstract}

\maketitle

\section{Introduction}

Special Relativity (SR) postulates Lorentz invariance as an exact symmetry of spacetime. It is at the basis of the Quantum Field Theory (QFT) description of the fundamental interactions, and has surpassed all experimental tests up to date (\cite{Kostelecky:2008ts,Long:2014swa,Kostelecky:2016pyx,Kostelecky:2016kkn}; see also the papers in~\cite{LectNotes702}). However, in a Quantum Gravity Theory (QGT) the usual concept of spacetime in SR would be replaced by a novel structure we are not able to imagine nowadays.

This has been an area of research in recent years. For example, in loop quantum gravity~\cite{Sahlmann:2010zf,Dupuis:2012yw}, such structure takes the form of a spin foam~\cite{Wheeler:1955zz,Rovelli:2002vp,Ng:2011rn,Perez:2012wv},   which can be interpreted as a ``quantum'' spacetime, and in causal set theory~\cite{Wallden:2013kka,Wallden:2010sh,Henson:2006kf} and string theory~\cite{Mukhi:2011zz,Aharony:1999ks,Dienes:1996du}, non-locality effects appear~\cite{Belenchia:2014fda,Belenchia:2015ake}. All these approaches differ completely from the notion of  Einstein's spacetime~\cite{Einstein1905}, which is constructed via  the exchange of light signals. Moreover, if there is a  ``quantum'' spacetime, the propagation of massless particles with different energies could vary, for example, through an energy dependent velocity, and Einstein's construction would lose sense. Furthermore, the description proposed by Einstein is not valid when non-locality effects arise.

Despite this complex structure being far from fully understood, a novel way of thinking arose not long ago: tiny effects in the propagation of high-energetic particles due to the quantum nature of spacetime could possibly be amplified by cosmological distances and produce observable consequences. For example, time delays in the arrival of photons with different energies could be observed, due to an energy-dependent velocity of propagation and the long distances travelled from the source to our telescopes~\cite{Amelino-Camelia1998,Vasileiou:2013vra,Ellis:2018lca}. 

Then, instead of considering a fundamental theory of quantum gravity, one can consider a down to top approach in which possible residual effects of a low-energy limit of a QGT could serve us as a guidance in the construction of such a theory. This idea is very promising in the current multi-messenger era we are actually living~\cite{Bartos2017}. A recent example of it is the detection of gravitational waves and gamma rays coming from the same binary neutron star merger,   GW170817~\cite{TheLIGOScientific:2017qsa} and GRB170817A~\cite{Goldstein:2017mmi}, respectively. 

In particular, Very High-Energy (VHE) gamma rays experiments, like the \emph{High Energy Stereoscopic System}  (H.E.S.S.) collaboration~\cite{HESS:2018zix}, have pointed out that the observed Extragalactic Background Light (EBL) seems to be lower than the one estimated by models such as the \emph{minimal EBL model}~\cite{Kneiske:2010}, i.e., the Universe seems to be more transparent to gamma rays than expected. This is known as  the \emph{pair production anomaly}~\cite{DeAngelis:2013jna}. Although these indications have somehow weakened lately \cite{Franceschini:2021wkr}, we note here that the physics of the Universe transparency is sensitive to modifications in the kinematics that could be related to quantum gravity phenomenology~\cite{Carmona:2020whi}. 
  
Possible deviations of special relativistic kinematics have been widely studied in the literature and can be separated into two completely different scenarios. In Lorentz Invariance Violation (LIV) scenarios~\cite{Colladay:1998fq,Kostelecky:2008ts}, the modification on the kinematics arises through a modified dispersion relation for particles. This scheme considers that Lorentz symmetry is violated at high energies, losing the relativistic invariance that characterizes SR, and then, causing a privileged observer to appear.  

In Deformed/Doubly Special Relativity (DSR)  theories~\cite{AmelinoCamelia:2008qg}, however, a relativity principle is still present. The kinematics of these theories is composed by a deformed dispersion relation, a deformed (non-additive) composition law for momenta, and, in order to have a relativity principle, there are Lorentz transformations making the two previous ingredients compatible with each other. It is important to note that the main ingredient of these theories is not the dispersion relation as in LIV scenarios, but the deformed composition law. In particular, a well-known example of DSR kinematics in which the dispersion relation is the one of special relativity is the classical basis of $\kappa$-Poincaré~\cite{Borowiec2010}, in which all the modification is introduced at the level of a deformed composition law.

Since both the kinematic ingredients and the theoretical implications are different in LIV and DSR, the phenomenological consequences are also different. For example, while an explicit formula of time delays (considering an expanding Universe) was derived for LIV scenarios~\cite{Ellis:2002in,Ellis:2005wr,RodriguezMartinez:2006ee}, some papers find such an effect for DSR theories~\cite{AmelinoCamelia:2012it,AmelinoCamelia:2011nt}, whilst other works have pointed out its possible absence~\cite{Carmona:2017oit,Carmona:2018xwm,Carmona:2019oph,Relancio:2020mpa}. As commented above, the main ingredient of DSR kinematics is the deformed composition law, and a deformation on the dispersion relation is not mandatory. This means that the new physical effects will appear in processes where two or more particles are involved, while the propagation of free particles might be unmodified. Therefore, the strong constraints on the high-energy scale imposed by time  delay experiments~\cite{Du:2020uev,Ackermann:2009aa,Wang:2016lne,Vasileiou:2015wja} might not be valid for DSR scenarios, and the high-energy scale parametrizing the new effects could be much lower than the Planck scale, e.g., of the order of TeV, and still be compatible with current experimental observations~\cite{Carmona:2020whi,Albalate:2018kcf}.

The propagation of VHE photons from astrophysical sources to our detectors has been previously analyzed within LIV kinematics. In this case, the modification of the dispersion relation does not only play a role through an energy dependent velocity of photons,  but also in the threshold energies of the interactions they can have during their travel, which certainly would affect their energy spectrum at Earth. The relevant interaction for the propagation of VHE photons is their collision with the low-energy photon backgrounds of the EBL or the Cosmic Microwave Background (CMB) to produce electron-positron pairs. In fact, a modification of the opacity  has been obtained within the LIV scenario~\cite{Martinez-Huerta:2020cut}, showing that the Universe could be more or less transparent depending on whether the modification of the dispersion relation leads to a subluminal or superluminal velocity. The threshold energy in LIV increases in the subluminal scenario, while it decreases for the superluminal case. 
It is common to consider that the only modification in the computation of the mean free path arises from the new value of the threshold energy, neglecting other possible LIV corrections on the cross section~\cite{Lang:2017wpe}. 
The LIV effects on the gamma ray flux,
which are already relevant for an energy scale that parametrizes the violation of the Lorentz invariance of the order of the Planck mass, have been studied in Ref.~\cite{Martinez-Huerta:2020cut}, where the model of~\cite{Dominguez:2010bv} for the EBL is used. The superluminal LIV case leads to a faster reduction of the gamma ray flux, which means that in this scenario the gamma rays interact more with the background than in SR.

The aim of this work is to study the transparency of the Universe to VHE photons in the scheme of DSR, or, generically speaking, in the framework of a Relativistic Deformation of the Kinematics (RDK): a modification of the kinematics of special relativity that is characterized by a non-trivial composition law of energy-momentum parametrized by a high-energy scale $\Lambda$, which is relativistic invariant, and reduces to special relativity in the $\Lambda\to\infty$ limit~\cite{Carmona:2019fwf,Carmona:2019vsh,Carmona:2021gbg}.

The present study is motivated by the recent observations of the \emph{Large High Altitude Air Shower Observatory} (LHAASO), which was able to report the detection of more than 530 photons with energies between $10^{14}\,\text{eV}$ and $1.4\cdot 10^{15}\,\text{eV}$ from 12 sources of ultra-high energy gamma rays within our galaxy, called \emph{PeVatrons}, with a statistical significance greater than 7 standard deviations~\cite{Cao2021}. The consequences of these results for LIV models have already been examined~\cite{Satunin:2021vfx,Chen:2021hen,Li:2021tcw,Li:2021cdz,LHAASO:2021opi}. The LHAASO results severely constrain the superluminal scenario, since pair-emission ($\gamma \to e^+ e^-$) and photon splitting ($\gamma\to 3\gamma$) processes force the energy scale characterizing the modification of the dispersion relation to be at least five orders of magnitude above the Planck energy $\approx1,22\cdot 10^{28}\,\text{eV}$ in the case of a linear correction~\cite{Li:2021tcw,LHAASO:2021opi}.
On the contrary, there are not such hard constraints on the subluminal LIV scenario, which enhances the transparency of the Universe to high-energy gamma rays.

In this work we carry out, for the first time, a similar analysis in the case of a relativistic deformed kinematics. The reason that the DSR scenario has not previously been considered for such a problem is due to the fact that usually, in the DSR literature, the high-energy scale deforming the kinematics is assumed to be of the order of the Planck energy, and in this case the effects of a relativistic kinematics would be completely irrelevant~\cite{Carmona:2020whi}. However, as mentioned above, the strong constraints on the high-energy scale obtained from the absence of observations of time delays for photons are not applicable with full generality in DSR. In the following, we will abandon the prejudice of a deformation scale of the order of the Planck scale and will consider how a RDK modifies the transparency to VHE photons and the constraints on the deformation scale that the LHAASO results impose.

In particular, we will focus our study on the interaction of high-energy gamma rays with the CMB photons, since they lead to the main contribution to the production of electron-positron pairs by photons with energies greater than  $10^{14}\,\text{eV}$~\cite{DeAngelis:2013jna}, which are in fact the ones in which we are interested because of the recent LHAASO results. This is the reason why one can neglect EBL photons in the calculation of the transparency of outer space at such ultra-high energies.

The structure of the paper is as follows. In Sec.~\ref{sec:DSR} we start by revising the computation of the mean free path in SR. We will then introduce a model of relativistic deformed kinematics and will compute the modification produced in the absorption of VHE photons, depending on the value of the high-energy scale, giving a detailed comparison between the mean free paths in the standard and the deformed kinematics. Finally, we conclude in Sec.~\ref{sec:conclusions}, contrasting our results with those obtained in the case of standard SR and LIV kinematics.

\section{Modification of the mean free path due to a relativistic deformed kinematics}
\label{sec:DSR}

In this section we start by reviewing the study, in special relativity, of the mean free path of VHE photons propagating in the cosmological electromagnetic background, showing the main computations that will allow us to generalize the result to DSR theories. To illustrate how the mean free path is modified due to a relativistic deformed kinematics, we will use a simple example of RDK that is characterized by the fact that the energy-momentum composition law contains only non-linear terms proportional to $1/\Lambda$, being $\Lambda$ the high-energy scale deforming the kinematics of SR. It can be seen~\cite{Carmona:2019vsh} that such a model is an example of what is known as $\kappa$-Poincaré kinematics~\cite{Lukierski:1991pn}. This example will allow us to deduce the modification of the mean free path in a particularly simple way.

\subsection{Computation of the photon mean free path in special relativity}
\label{fondos}

VHE photons ($\gamma$) interact with CMB photons ($\gamma_{b}$) producing electron-positron pairs, $\gamma + \gamma_b \to e^+ + e^-$, thereby reducing their flux in their way towards our telescopes. 

The minimum energy needed by the background photon, the threshold energy $\epsth$, varies as a function of the energy of the incident photon, $E$, and the angle formed by both photons, $\theta$. This energy must be such that the squared center of mass energy, $s=2 E \varepsilon (1-\cos\theta)$, reaches its lowest possible value, $4m_e^2$. Then, the threshold energy is  
\begin{equation}
    \epsth\,=\, \frac{2\,m_e^2}{E\,(1-\cos{\theta})}\,.
    \label{umbral}
\end{equation}

The cross section of the pair production, the well-known Breit-Wheeler cross section, is given by~\cite{DeAngelis:2013jna,Gould:1967zzb,Nikishov1961}:
\begin{equation}
    \sigma_{\gamma\gamma}(E,\varepsilon,\theta)\,=\,\frac{2\pi \alpha^2}{3m_e^2} \, W(\beta)\,,
    \label{seccioneficaz}
\end{equation}
being 
\begin{equation}
     W(\beta)\,=\, (1-\beta^2)\,\left[\,2\beta (\beta^2-2)+(3-\beta^4)\,\ln{\frac{1+\beta}{1-\beta}}\,\right]\,,
     \label{W}
\end{equation}
with $\alpha \approx 1/137$, and $\beta$ the velocity of the electron and positron in the center of mass reference frame,
\begin{equation}
    \beta(\varepsilon,E,\theta)\,=\,\sqrt{1-\frac{2\,m_e^2}{\varepsilon\,E\,(1-\cos{\theta})}}\,=\,\sqrt{1-\frac{4m_e^2}{s}}\,.
    \label{beta}
\end{equation}
Then, the cross section can be expressed as a function of the relativistic invariant $s$. 

The interaction between VHE photons and background photons leads to a suppression of the intrinsic flux of photons, $\Phi$. The observed flux of photons is given by 
\begin{equation}
   \Phi_{\text{obs}}(E,z_s)\,=\, \exp{(-\tau_{\gamma}(E, z_s))}\,\Phi(E(z_s))\,,
    \label{flujo}
\end{equation}
where $\tau_{\gamma}(E, z_s)$ is the optical depth, $z_s$ is the redshift of the source, and $E(z_s)=E (1+z_s)$ is the energy of the photons emitted by the source. 

The expression for the optical depth due to the production of electron-positron pairs in the interaction of VHE photons and background photons is~\cite{Gould:1967zzb,Nikishov1961} 
\begin{equation}
      \tau_{\gamma}(E,z_s)\,=\, \int^{z_s}_0 dz\, \frac{d l(z)}{dz}\int^1_{-1} d(\cos{\theta})\,\frac{1-\cos{\theta}}{2} \int^{\infty}_{\epsth} d\varepsilon\,n_{\gamma}(\varepsilon,z)\,\sigma_{\gamma\gamma}(E(z),\varepsilon,\theta)\,,
    \label{tau}
\end{equation}
where $\sigma_{\gamma\gamma}(E(z),\varepsilon,\theta)$ is the pair-production cross section, $E(z)$ is $E(1+z)$, $dl(z)/dz$ is the distance traveled by a photon per unit redshift at redshift $z$, which depends on the cosmological parameters~\cite{DeAngelis:2013jna},
and $n_{\gamma}(\varepsilon,z)$ is the spectral density of background photons, which in general depends on $z$ and is the sum of the different background contributions of low-energy photons: EBL, CMB and Radio Background (RB).

If we focus on the local Universe, we can disregard the cosmological effect of the expansion and use the photon mean free path, $\lambda_\gamma(E)$, instead of the optical depth in order to quantify the transparency of the Universe. The survival probability
$P_{\gamma\rightarrow\gamma}(E,z_s)=\exp(-\tau_\gamma(E,z_s))$ for photons of energy $E$ from a source at a redshift $z_s$ can be approximated in the case of near sources by 
$P_{\gamma\rightarrow\gamma}(E,D)\approx \exp(-D/\lambda_\gamma(E))$,
where  $D$ is the distance from the source. Lower values of the optical depth and greater mean free paths correspond to a lower absorption of the VHE photons by the background. The mean free path for the VHE photon is then given by
\begin{equation}
     \frac{1}{\lambda_\gamma(E)}  \,=\, {\int^1_{-1} d(\cos{\theta})\frac{1-\cos{\theta}}{2}\int^{\infty}_{\epsth} d\varepsilon \, n_{\gamma}(\varepsilon)\,\sigma_{\gamma\gamma}(E,\varepsilon,\theta)}\,,
     \label{lambdaSR}
\end{equation}
where $n_{\gamma}(\varepsilon)$ is the local spectral density of background photons, $n_\gamma(\varepsilon)=\lim_{z\to 0}n_\gamma(\varepsilon,z)$.  

\begin{figure}[tbp]
    \centering
    \includegraphics[scale=0.32]{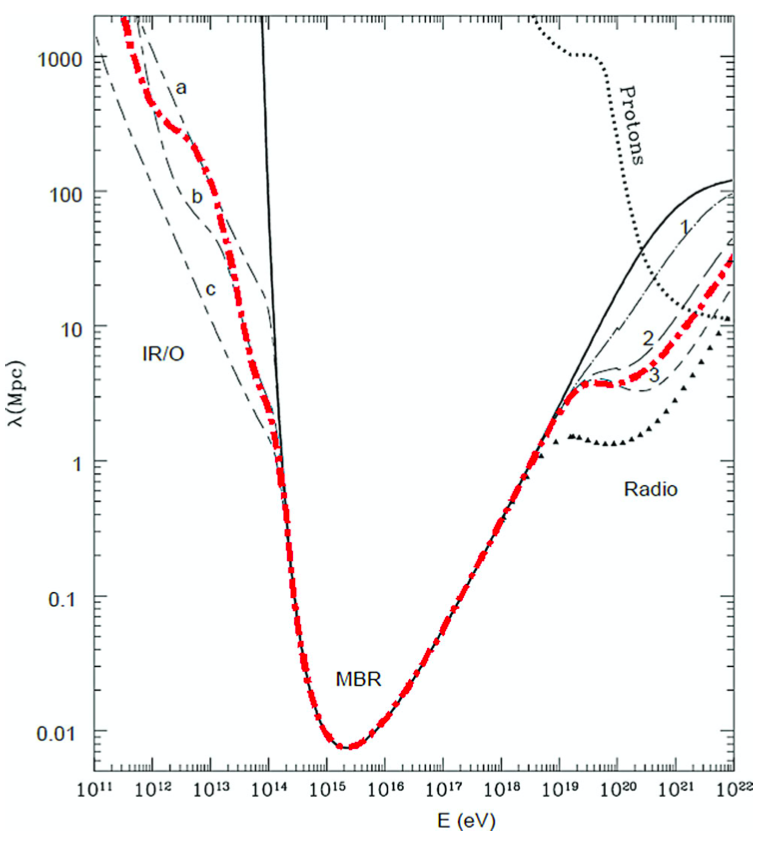}
    \caption{In red, the mean free path $\lambda$ of photons for pair production computed by Ref.~\cite{DeAngelis:2013jna} considering all backgrounds. The black solid line corresponds to the CMB (Microwave Background Radiation or MBR), black lines labelled by a, b, c represent different EBL models (Infrared and Optical Backgrounds or IR/O), black lines labelled by 1, 2, 3 represent a model of the Radio background with different low-frequency cutoffs, and the black triangles correspond to the radio background under the assumption that it is completely extragalactic. For an energy around $2\,\text{PeV}$, the mean free path reaches a minimum. On the upper right corner the mean free path of protons is shown for comparison. Figure extracted from ``Transparency of the Universe to gamma rays'', figure 2~\cite{DeAngelis:2013jna}.}
    \label{fig:figura2}
\end{figure}

Figure~\ref{fig:figura2}, taken from Ref.~\cite{DeAngelis:2013jna}, shows the mean free path as a function of the energy of the incident photon. One can observe that for the range of energies where the CMB dominates, $\lambda_\gamma$ is much smaller than for the energy range where the dominant background is the EBL or RB. It has a minimum at around $2\,\text{PeV}$ and increases at high energies due to a decreasing in the cross section, leading to a greater transparency at energies where the dominant background is the RB.

It is important to note that the recent observations by LHAASO~\cite{Cao2021} involve energies very close to the minimum of the mean free path. For the energies in this range of interest (greater than $10^{14}\,\text{eV}$), we can compute the mean free path by using only the contribution of the CMB photons, which dominates in this region over the EBL or radio background until energies above $10^{18}\,\text{eV}$.
Since the CMB can be modeled by a black body emitter, its spectral density is known. This has a clear advantage over other backgrounds, like the EBL one, for which the spectral density must be obtained experimentally, having then uncertainties~\cite{Dominguez2010}. 

For $z\ll1$, the spectral density of CMB photons is given by~\cite{Fixsen:1996nj}
\begin{equation}
    n_{\gamma} (\varepsilon)\,=\,(\varepsilon/\pi)^2(e^{\varepsilon/kT_0}-1)^{-1}\,.
\end{equation}

In order to obtain the mean free path  $\lambda_{\gamma}$, we start from Eq.~\eqref{lambdaSR}, which is a double integral on the angle $\theta$ formed by both photons and the energy $\varepsilon$ of the background photon. For the sake of simplicity for further computations, we can express the double integral as a function of the squared  center of mass energy  by the following change of variables:
\begin{equation}
   \frac{1- \cos{\theta}}{2} \,=\, \frac{s}{4E\varepsilon}\qquad \implies \qquad d\left(\frac{1- \cos{\theta}}{2}\right)\,=\,\frac{ds}{4E\varepsilon} \,.
\end{equation}
The limits of integration are then:
\begin{equation}
    4 m_e^2\,\leq \,s\,< \,\infty\,,\quad\quad\frac{s}{4E}\,\leq \,\varepsilon\,<\, \infty\,.
\end{equation}
Therefore, Eq.~\eqref{lambdaSR} can be written as
\begin{equation}
     \frac{1}{\lambda_\gamma(E)}\,=\, 2\int^1_{0} d\left(\frac{1- \cos{\theta}}{2}\right)\frac{1-\cos{\theta}}{2} \int^{\infty}_{\frac{m_e^2/E}{(1-\cos{\theta})/2}} d\varepsilon \, n_{\gamma}(\varepsilon)\sigma_{\gamma\gamma}(E,\varepsilon,\theta)\\
     \,=\, \frac{1}{8E^2}\int^{\infty}_{4 m_e^2} ds\,s\,\sigma_{\gamma\gamma}(s)\int^{\infty}_{s/4E} d\varepsilon\,\frac{n_{\gamma}(\varepsilon)}{\varepsilon^2}\,.
     \label{lambda_completo}
\end{equation}
We can use the following dimensionless variables
\begin{equation}
    \Bar{s}\,=\,\frac{s}{4m_e^2}\,, \qquad \Bar{\varepsilon}\,=\,\frac{\varepsilon}{kT}\,,\qquad \Bar{E}\,=\,\frac{E}{m_e^2/(kT)}\,\approx\,\frac{E}{1,1\, \text{PeV}},
    \label{variablesnuevas}
\end{equation}
where $kT=2,35\cdot10^{-4}\,\text{eV}$, and $m_e^2=2,61\cdot10^{11}\,\text{eV}^2$. The limits of integration in the new dimensionless variables are:
\begin{equation}
    1\,\leq\, \Bar{s}\,<\, \infty\,,\quad\quad\frac{\Bar{s}}{\Bar{E}}\,\leq\, \Bar{\varepsilon}\,<\, \infty.
\end{equation}
This allows us to express the mean free path in a compact fashion,
\begin{equation}
\centering
    \frac{1}{\lambda_\gamma(E)} \,=\,\frac{4\alpha^2 (kT)^3}{3m_e^2 \pi}\, \frac{1}{\Bar{E}^2}\int^{\infty}_{1} d\Bar{s}\,\Bar{s}\,\overline{W}(\Bar{s})\int^{\infty}_{\Bar{s}/\Bar{E}} d\Bar{\varepsilon}\,(e^{\Bar{\varepsilon}}-1)^{-1}\,,
    \label{lambdabarra}
\end{equation}
where $\overline{W}(\Bar{s})\doteq W(\beta(\Bar{s}))$ and $\beta(\Bar{s})=\sqrt{1-1/\Bar{s}}$.

When computing numerically the previous equation, we obtain the same results as indicated by the black solid line in Fig.~\ref{fig:figura2}, except for energies  greater than  $E=10^{20}\,\text{eV}$. This is due to the fact that in~\cite{DeAngelis:2013jna} additional effects, such as  double pair production, which become important at these energies, are taken into account~\cite{Demidov2008,Protheroe1995}. These complications will not be relevant in our case, since we are interested in the phenomenology that present and near future experiments can explore, which lies some orders of magnitude below those energies. The result of Ref.~\cite{Anchordoqui2002}, which does not include additional effects to pair production, coincide with our result and provides an additional check of our computations.

\subsection{Simple model of relativistic deformed kinematics}

After this review about how the mean free path of photons is obtained in special relativity, we want to extend this computation to the case of doubly special relativity. As commented in the Introduction, the main ingredient of DSR kinematics is a deformed composition law. In this work, for the sake of simplicity, we will use the  simple basis of $\kappa$-Poincaré kinematics obtained in~\cite{Carmona:2019vsh}. The deformed composition law is derived from the requirements of linearity as a function of the four-momentum of each particle, isotropy, associativity, and asymmetry under exchange of the two four-momenta.

The isotropy implies that rotations are not affected by the deformation, a condition satisfied by most of the examples of a nonlinear implementation of the Lorentz invariance that have been considered, either from an algebraic or a geometric perspective. Associativity makes trivial the extension of the composition law from a two-particle to a multiparticle system  and also makes the deformation of the kinematics compatible with a simple implementation of locality in a generalized spacetime~\cite{Carmona:2019vsh}. An asymmetry under exchange of the two four-momenta  guarantees that one has a real deformation of SR instead of a simple change of momentum variables. The requirement of linearity as a function of the four-momentum of each particle is a consequence of a choice of momentum variables; in fact, it can be shown that the following change of momentum variables  $k_\mu=(k_0,\vec{k}) \to k'_\mu=(k'_0,\vec{k'}) $ 
\be
k_i \,=\, k'_i\,, \quad\quad\quad (1 - k_0/\Lambda) \,=\, e^{- k'_0/\Lambda}\,,
\ee
relates this composition law to the composition law corresponding to the coproduct of $\kappa$-Poincaré in the bicrossproduct basis (primed variables)~\cite{Carmona:2019vsh}. Here, $\Lambda$ denotes the high-energy scale deforming the special relativistic kinematics. The role that different choices of momentum  variables play in a relativistic deformed kinematics is still an open question. 

The deformed composition law of two momenta $p$ and $q$ satisfying the previous requirements is
\be
\left(p\oplus q\right)_\mu \,=\, p_{\mu} + \left(1 + p_0/\Lambda\right) \,q_{\mu}\,,
\label{eq:dcl1}
\ee 
where $p_0$ denotes the zero component (energy) of the four-momentum $p_\mu$. It is easy to see that for $\Lambda$ going to infinity, one recovers the addition law of SR, viz. the sum. 

The compatibility of the deformed conservation of energy-momentum, corresponding to the composition law (\ref{eq:dcl1}), with the relativity principle leads to identify a non-linear implementation of Lorentz transformations of a two-particle system and also of the momentum of one particle. Invariance under the nonlinear Lorentz transformation of a four-momentum variable allows one to derive the deformed dispersion relation,
\be
\frac{k_0^2-\vec{k}^2} {1+k_0/\Lambda}\,=\,m^2\, ,
\label{casimir}
\ee
which is indeed compatible with the deformed composition law (\ref{eq:dcl1}). Again, by taking the limit $\Lambda\to \infty$ one finds the quadratic expression of SR. This is all we need to see how the deformation affects the kinematics of the electron-positron production in the interaction of a high-energy photon with a low-energy CMB photon.

\subsection{Calculation of the mean free path in the relativistic deformed kinematics}

The derivation of the expression~\eqref{lambdabarra} for the mean free path in SR is based on the expression for the squared center of mass energy $s=(p+q)^2$, where $p$ and $q$ are either the momenta of the two photons in the initial state, or those of the electron and positron in the final state. In the model of relativistic deformed kinematics considered in this work,
the squared center of mass energy is defined by the deformed dispersion relation~\eqref{casimir} and the  total momentum~\eqref{eq:dcl1},
\begin{equation}
    \Tilde{s}\,=\,\frac{(p \oplus q)^2}{1+ (p \oplus q)_0/\Lambda}\,,
    \label{s deformado}
\end{equation}
where we use the tilde notation to distinguish it from the usual relativistic invariant of SR.

In the case of the collision of two photons with energies $E_1$ and $E_2$, one gets the following expression
\begin{equation}
    \Tilde{s}\,=\,\frac{2\,E_1\,E_2\,(1-\cos{\theta})}{1+E_2/\Lambda}\,.
\end{equation}
Since the deformed composition law is not symmetric, we find that there are two possible orderings of the momenta. In the particular case considered here, either the momentum of the VHE photon is composed with the momentum of the background one, or vice-versa. Then, we have two different channels through which the process can  take place. 

The two different invariants  are 
\begin{equation}
    \Tilde{s}_1\,=\,\frac{2\,E\,\varepsilon\,(1-\cos{\theta})}{1+\varepsilon/\Lambda}\,\approx \,2\,E\,\varepsilon\,(1-\cos{\theta})\, ,
    \label{s'1}
\end{equation}
\begin{equation}
    \Tilde{s}_2\,=\,\frac{2\,\varepsilon\,E\,(1-\cos{\theta})}{1+E/\Lambda}\,,
\end{equation}
where in the first invariant we have used that $\varepsilon/\Lambda \rightarrow0$, since the energy of the background photon is very small in comparison with the high-energy scale. Then, the first invariant~\eqref{s'1} coincides with the special relativistic one. 

Due to our particular and simple choice of the deformed kinematics, we can express the new invariant $\Tilde{s}_2$ as the same one of SR with a different energy of the VHE photon, $E'$:
\begin{equation}
    \Tilde{s}_2\,=\,\frac{2\,\varepsilon\,E\,(1-\cos{\theta})}{1+ E/\Lambda}\,=\,2\,\varepsilon\,E'\,(1-\cos{\theta})\,,
    \label{s'2}
\end{equation}
where
\begin{equation}
    E'\,=\,\frac{E}{1+ E/\Lambda}\,.
\end{equation}
The functions \eqref{s'1} and \eqref{s'2} are the same function evaluated at different energies:
\begin{equation}
      \Tilde{s}_1\,=\,s(E)=2\,\varepsilon\,E\,(1-\cos{\theta})\,,\quad\Tilde{s}_2\,=\, s(E')=2\,\varepsilon\,E'\,(1-\cos{\theta}) \,.
    \label{eq:relation-invariants}
\end{equation}

Once we have identified the effect of the deformed kinematics on the relativistic invariant $s$, we need the generalization of the Breit-Wheeler cross section $\sigma_{\gamma\gamma}(s)$ of pair production. In contrast to what happens in special relativity~\cite{Maggiore:2005qv}, in the case of a LIV scenario a cross section is not a relativistic invariant, but it can be explicitly computed in a specific system of reference in frameworks such as the Lorentz-violating Standard Model extension~\cite{Colladay:2001wk}. We do not have at present a well defined dynamical framework incorporating a relativistic deformed kinematics, but, assuming its existence, the compatibility of the deformation with relativistic invariance will imply that the deformed cross section in such a theory will be a function of the deformed invariant (\ref{s deformado}) with a possible additional explicit dependence on the scale of deformation $\Lambda$. But, taking into account that $(\tilde{s}/\Lambda^2)\ll 1$, one has that $\tilde{\sigma}_{\gamma\gamma}(\tilde{s},\tilde{s}/\Lambda^2)\approx\tilde{\sigma}_{\gamma\gamma}(\tilde{s},0)\doteq\tilde{\sigma}_{\gamma\gamma}(\tilde{s})$,
and, since when $\tilde{s}\to s$ the deformed cross section should approach the SR cross section, one gets that $\tilde{\sigma}_{\gamma\gamma}\approx \sigma_{\gamma\gamma}$. We will use this approximation (as in Refs.~\cite{Albalate:2018kcf,Carmona:2021pxw}) to compute the effect of the deformation in the dynamics producing the electron-positron pair.

Together with the simple relation~\eqref{eq:relation-invariants} between the deformed and the SR invariants, we have to see which is the lower limit in the integral over $\tilde{s}$ in the evaluation of the modified mean free path. Being a relativistic invariant, it can be calculated in any reference frame. In particular we can consider a reference frame where the total momentum of the two photons in the initial state (and then also the total momentum of the electron-positron pair in the final state) is zero. In the case of SR one can have the two photons with the same energy $E=\epsilon=m_e$ in this reference frame and the electron and positron at rest so that $s_\text{min}=4m_e^2$. In the case of a relativistic deformed kinematics one will have  $\tilde{s}_{\text{min}}=4m_e^2(1+\mathcal{O}(m_e/\Lambda))$,  since the correction should be inversely proportional to $\Lambda$.  Then, at the threshold of the electron-positron pair production, the modifications due to the deformation of the relativistic kinematics can be neglected, and  we can approximate the lower limit in  the integration over $\tilde{s}$ by $4 m_e^2$.   

\begin{figure}[tbp]
    \centering
    \begin{minipage}{0.49\textwidth}
        \centering
        \includegraphics[width=\textwidth]{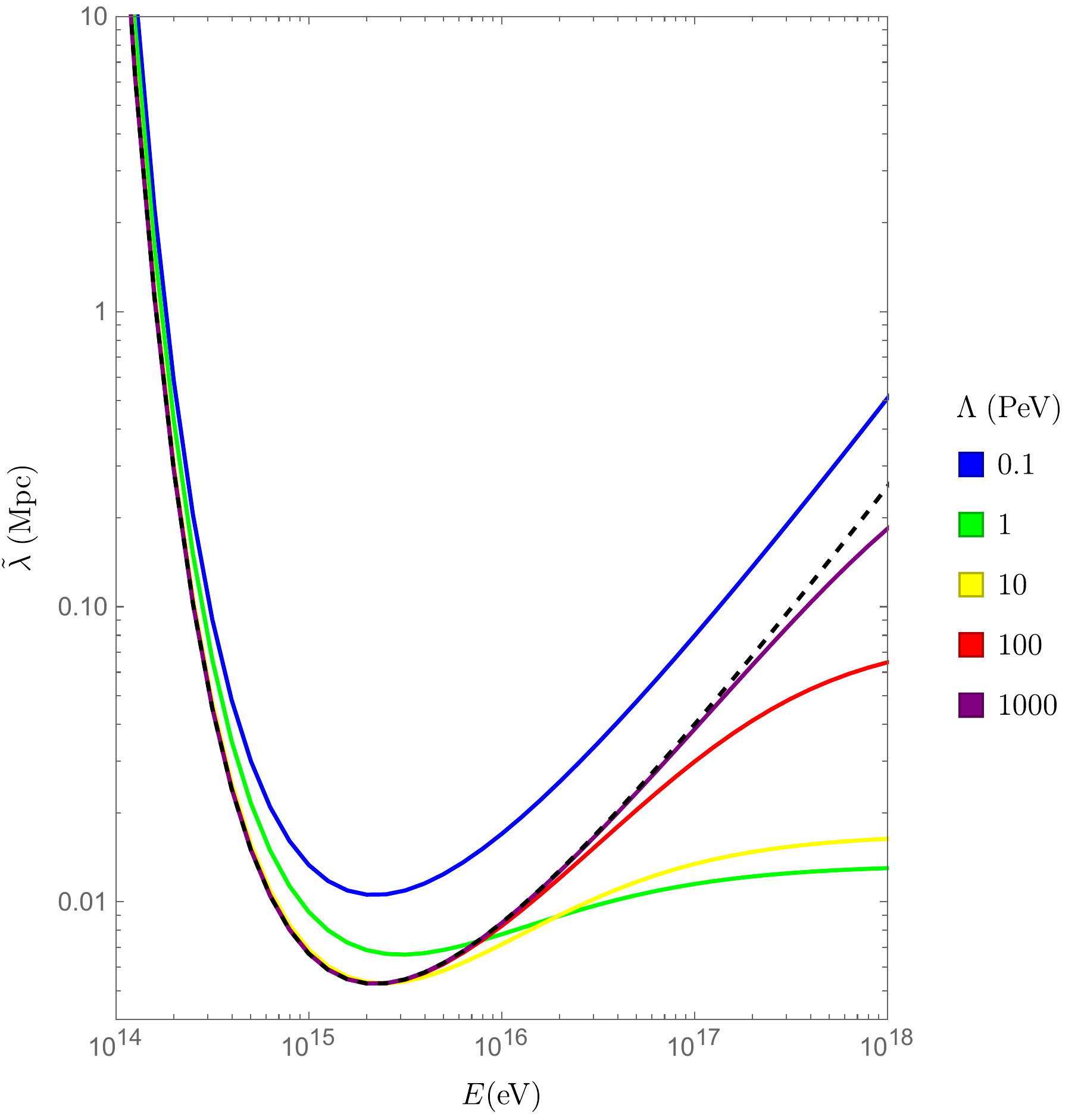}
        \caption{Mean free path for photons as a function of its energy in the range where the CMB is the relevant background. The black dashed curve corresponds to the mean free path in SR, and the colored solid curves represent the result in RDK for different orders of magnitude of the scale $\Lambda$.}
        \label{fig:figura5}
    \end{minipage}%
    \hfill
    \begin{minipage}{0.49\textwidth}
        \centering
        \includegraphics[width=\textwidth]{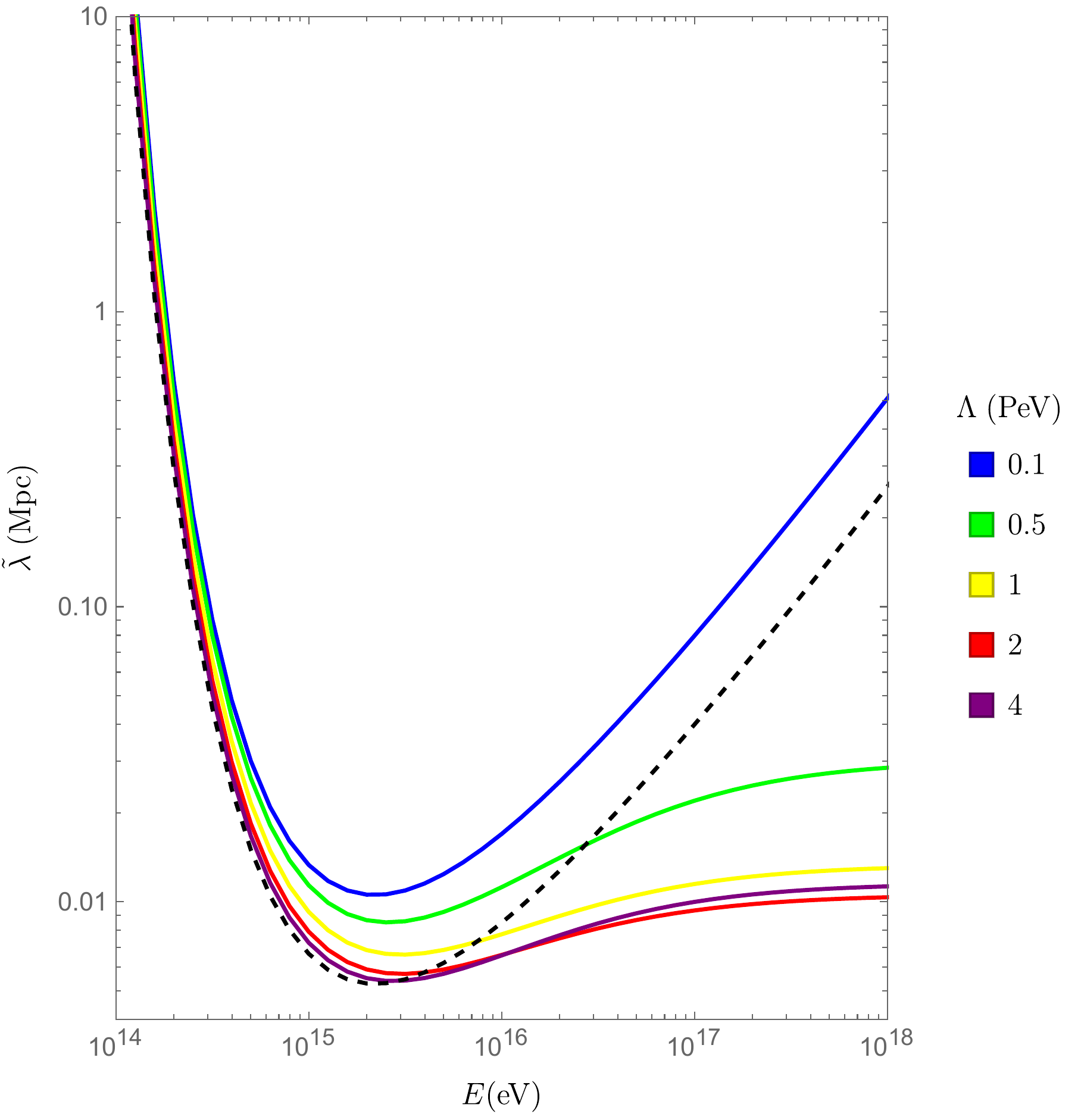}
        \caption{Mean free path for photons as a function of its energy. The black dashed curve corresponds to the mean free path in SR, and the colored solid curves represent the result in RDK for different values of the scale $\Lambda$ near the end of the observed LHAASO spectrum (PeV scale).}
        \label{fig:figura8}
    \end{minipage}
\end{figure}

Finally, lacking as mentioned a dynamical framework compatible with a relativistic deformed kinematics, we need to make an assumption on the probability of the two channels, corresponding to the two initial states of the photons with a different total momentum, through which the process can take place. The simplest hypothesis is to consider that both of them are equally probable, as it has been previously considered in several works~\cite{AmelinoCamelia:2001fd,Freidel:2013rra}. Then, the inverse of the mean free path will be given by the average of the inverse of the mean free paths calculated from~\eqref{lambda_completo} using each invariant $\Tilde{s}_1$ and $\Tilde{s}_2$,
\begin{equation}
\frac{1}{\Tilde{\lambda}(E)}\,=\,\frac{1}{2}\,\left(\frac{1}{\tilde{\lambda}_1(E)}+\frac{1}{\tilde{\lambda}_2(E)}\right)\,=\,\frac{1}{2}\,\left(\frac{1}{\lambda(E)}+\frac{1}{\lambda(E')}\right)\,.
    \label{lambda'}
\end{equation}
We conclude that the modified mean free path can be read from the one of SR through Eq.~\eqref{lambda'}, avoiding a different computation. This is the key ingredient for the results presented in this work. Fig.~\ref{fig:figura5} is obtained by applying Eq.~\eqref{lambda'}  in a range of energies for the high-energy photon where the CMB dominates over other backgrounds: from $10^{14}$ to $10^{18}$\,eV, for different orders of magnitude of the high-energy scale $\Lambda$. In Fig.~\ref{fig:figura8}, we focus on the behavior of the mean free path  for values of the scale $\Lambda$ for which one would have observable effects in LHAASO and other near future experiments.

It is important to remark that the derivation of the relation between the RDK and SR mean free paths in Eq.~\eqref{lambda'} does not depend on the photon background distribution $n_\gamma(\epsilon)$, and then it will be valid beyond the energy range where those mean free paths can be approximated by considering only the CMB component.
 
\subsection{Study of the mean free path curves and the expected observed flux}

Let us study in more detail the curves of the deformed mean free path at Fig.~\ref{fig:figura8}. Let us first understand the behavior at energies $E\gg \Lambda$, i.e., the right end of the graph. We should notice that at energies much larger than the scale of energy parametrizing the deformation of the kinematics one has
  \begin{equation}
      \frac{1}{\tilde{\lambda}(E)} \approx \frac{1}{2} \left(\frac{1}{\lambda(E)}+\frac{1}{\lambda(\Lambda)}\right)\,. 
  \end{equation}
For the curves such that $\lambda(E)\gg\lambda(\Lambda)$, the deformed mean free path approaches to a constant value $\tilde\lambda(E)\rightarrow 2\lambda(\Lambda)$; this is the behavior of the curves from $\Lambda=0.5\, \text{PeV}$ to $\Lambda= 4 \, \text{PeV}$. What is more, as the mean free path of SR, $\lambda(E)$, has a minimum at $E_0\approx 2 \, \text{PeV}$, the asymptotic value $2\lambda(\Lambda)$, as function of $\Lambda$, will decrease until $\Lambda\approx 2\, \text{PeV}$ and then it will start increasing; that is the reason why the curve of $\Lambda =4 \, \text{PeV}$ ends at higher values than the curve of $\Lambda=2 \, \text{PeV}$. For $\lambda(E)\ll\lambda(\Lambda)$, the deformed mean free path tends to $\tilde\lambda(E)\rightarrow 2\lambda(E)$; this is the behavior one can see in the curve of $\Lambda =0.1\, \text{PeV}$.

Another important characteristic of Fig.~\ref{fig:figura8} is the crossing point between the deformed mean free path and the one of SR, because it will distinguish the energies where the deformation of the kinematics will produce a larger or smaller mean free path than that of SR. The energy at which the mean free path curves of special relativity and the chosen relativistic deformed kinematics intersect, $E^*(\Lambda)$, satisfies
\begin{equation}
  \frac{1}{\lambda(E^*)}\,=\,\frac{1}{\Tilde{\lambda}(E^*)}=\frac{1}{2}\,\left(\frac{1}{\lambda(E^*)}+\frac{1}{\lambda(E^{*'})}\right)\,.
\end{equation}
so that
\begin{equation}
   \lambda(E^*)\,=\,\lambda(E^{*'})\,,\quad \text{where}\,\quad E^{*'}\,=\,\frac{E^*}{1+ E^*/\Lambda}\,.
   \label{corte}
\end{equation}

\begin{figure}[tbp]
    \centering
    \includegraphics[scale=0.5]{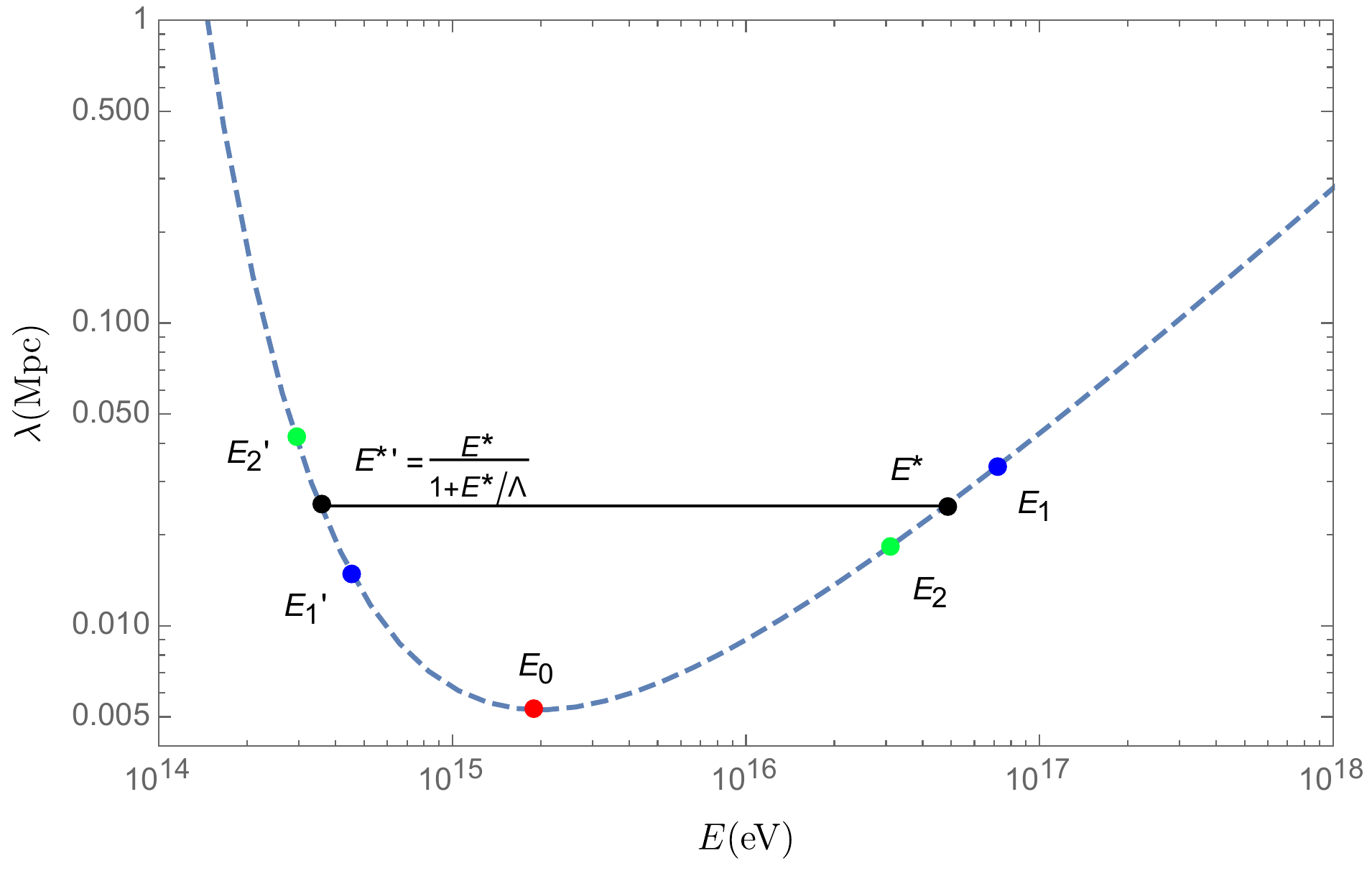} 
    \caption{Graphical representation of the position of $E^*(\Lambda)$ and $E^{*'}(\Lambda)$ as defined in Eq.~\eqref{corte} in the mean free path curve $\lambda(E)$. The red point is the minimum value of $\lambda(E)$; black points correspond to two different energies, $E^{*}(\Lambda)$ and $E^{*'}(\Lambda)$, which for a particular value of $\Lambda$ lead to the same mean free path $\lambda$; blue and green points corresponds to greater and smaller energies than $E^{*}(\Lambda)$.}
    \label{fig:figura6}
\end{figure}

For Eq.~\eqref{corte} to be satisfied, $E^*$  and $E^{*'}$ must be, respectively, to the right and left sides of the minimum of $\lambda(E)$, denoted as $E_0$ in Figure~\ref{fig:figura6}. For a given energy $E$, in order to know if in the RDK scenario the Universe is going to be more or less transparent than in SR, we have to compare $E$ with $E^*(\Lambda)$. We have two cases:

\begin{enumerate}
    
\item $E>E^*(\Lambda)$. Choosing an energy ($E_1$ in Figure~\ref{fig:figura6}) above  $E^*(\Lambda)$, $E_1'$ is greater than  $E^{*'}(\Lambda)$ and then
    \begin{equation}
        \lambda(E_1)\,>\,\lambda(E'_1) \,\implies\, \Tilde{\lambda}(E_1)\,<\,\lambda(E_1)\,.
        \label{46}
    \end{equation}
For energies greater than  $E^*(\Lambda)$, the mean free path in RDK is smaller than the one of SR, i.e., the Universe is less transparent in RDK than in SR.   
    
\item $E<E^*(\Lambda)$. Choosing an energy ($E_2$ in Figure~\ref{fig:figura6}) below $E^*(\Lambda)$, $E_2'$ is smaller than $E^{*'}(\Lambda)$ and then
    \begin{equation}
        \lambda(E_2)\,<\,\lambda(E'_2) \,\implies\, \Tilde{\lambda}(E_2)\,>\,\lambda(E_2).
        \label{47}
    \end{equation}
For energies smaller than $E^*(\Lambda)$, the mean free path is greater in RDK than in SR, so the Universe is more transparent in RDK than in SR.  
    
\end{enumerate}

\begin{figure}[tbp]
    \centering
    \begin{minipage}{0.49\textwidth}
        \centering
        \includegraphics[scale=0.48]{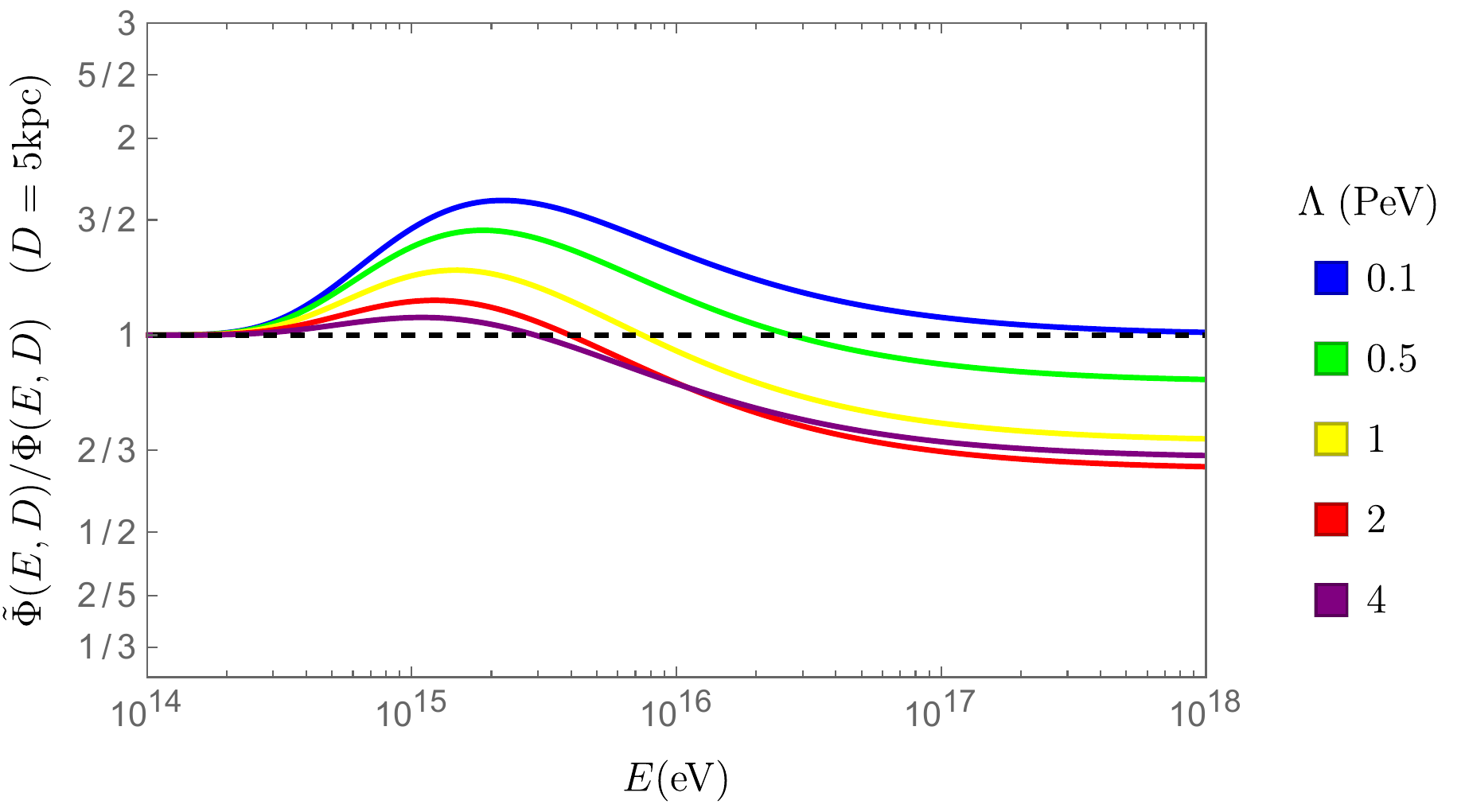}
    \end{minipage}%
    \hfill
    \begin{minipage}{0.49\textwidth}
        \centering
        \includegraphics[scale=0.48]{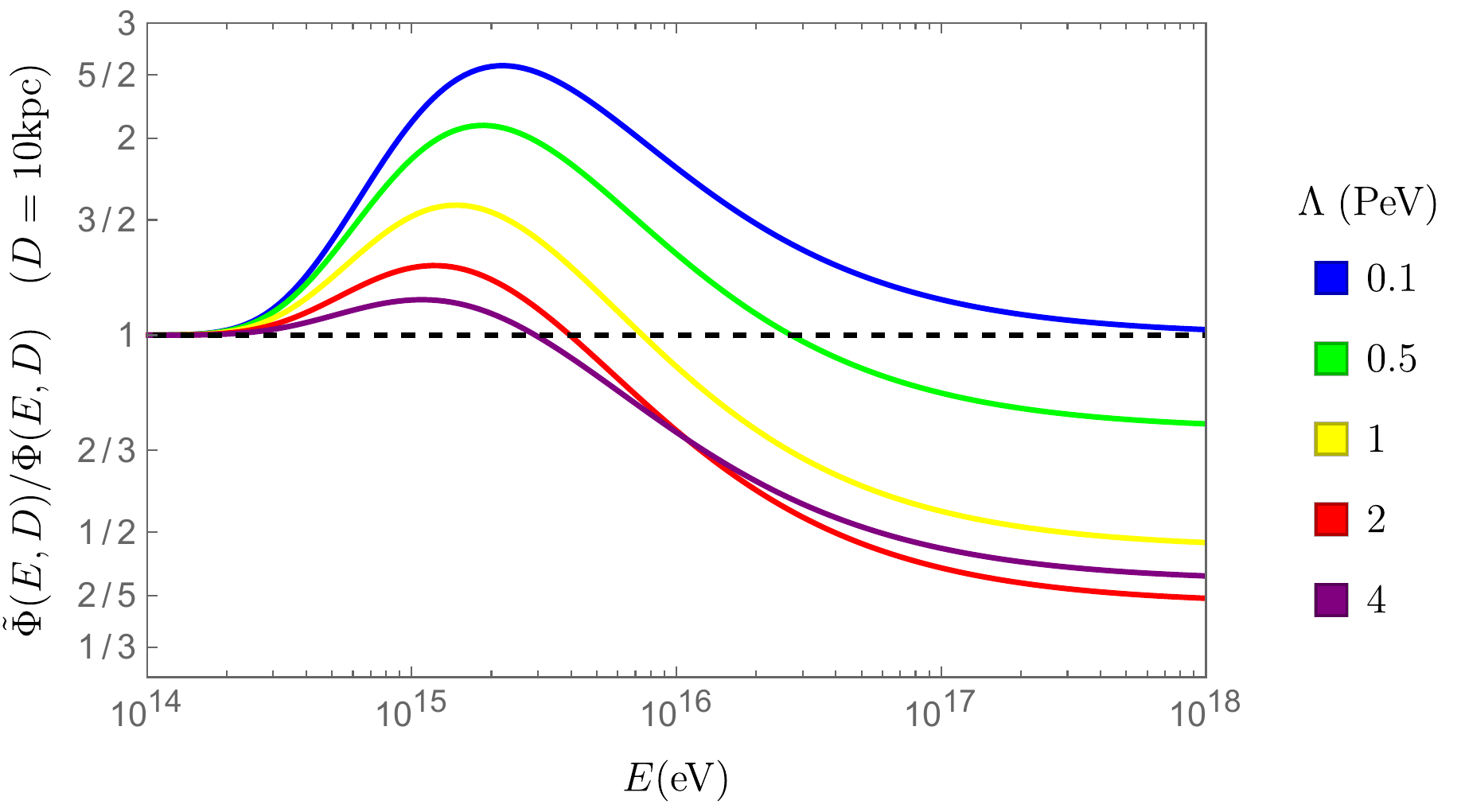}
    \end{minipage}
    \caption{Ratio between the observed fluxes of high-energy gamma rays in the deformed kinematics and special relativity cases for the values of $\Lambda$ considered in Fig. \ref{fig:figura8}, corresponding to sources at a distance $D=5\,$kpc (left) and $D=10\,$kpc (right). Note that both axes use logarithmic scale, so that the vertical separation between the curve and the dashed horizontal line is the same when the deformed flux increases or decreases with respect to the standard case by the same factor (e.g., the distance between 1 and 2 in the vertical axis is the same as between $1/2$ and 1).}
    \label{fig:figura11}
\end{figure}

This behavior can be clearly seen in Fig.~\ref{fig:figura8}, where one can also check that the crossing points of the mean free path of SR and RDK happen for greater energies as the deformation scale is smaller, and always for energies greater than the minimum of $\lambda(E)$.

The crossing points we have just found will also distinguish, for each value of the scale $\Lambda$, where we expect a larger or smaller observed flux of gamma rays with respect to the case of SR. We have plotted in Fig.~\ref{fig:figura11} the ratio between the observed fluxes of high-energy gamma rays coming from a galactic source at distance $D$ for the case of the deformed kinematics, $\tilde\Phi_\text{obs}(E,D)$, and of special relativity, $\Phi_\text{obs}(E,D)$, for the values of $\Lambda$ showed in Fig.~\ref{fig:figura8} and two different values of $D$. We can take this as a measure of the sensitivity to the deformed kinematics in the detection of high-energy photons from within our galaxy; the different values of $D$ shown in the figure are examples of the role of the uncertainty in the location or identification of the sources.

We observe in the figures that, for every value of $\Lambda$, there is an increase of the observed flux in the deformed case with respect to the non-deformed case at energies close to the minimum of the mean free path curve, $E_0$, which goes down when the energy approaches $E^*(\Lambda)$ (at which the curve crosses the dashed horizontal line), and a decrease in the deformed flux with respect to the standard one for energies above $E^*(\Lambda)$, which gets more pronounced at higher energies. Experiments would then observe an anomalously high transparency at energies $E\lesssim E^*(\Lambda)$ and an anomalously low transparency at energies $E\gtrsim E^*(\Lambda)$, which would be more evident from sources at large distances. It 
is however important to notice that, even if larger distances would make the anomalies more manifest, the sources cannot be much farther than the mean free path of the photons in order to have an observable flux at the detector.

\section{Conclusions}
\label{sec:conclusions}

In this work we have explored the possible phenomenological consequences of a relativistic deformed kinematics on the propagation of high-energy gamma rays in the local Universe.    
It is important to note that, in order to identify effects of new physics, due to a deformation of SR for example, the full knowledge of the sources and their emission would be mandatory. Indeed, we do not have such information and we only have different astrophysical models which determine the intrinsic flux of photons with uncertainties. This is a strong limitation when trying to get constraints to new physics beyond SR or to favor it as an alternative scenario.  

We have shown that the spectrum of very high-energy photons can differ appreciably in the case of DSR with respect to the expected spectrum in the special relativistic scenario for energies above  $\approx 2\times 10^{14}\,\text{eV}$ if the high-energy scale of deformation $\Lambda$ is of the order of the PeV scale.
Thanks to experiments like LHAASO, which recently detected photons around this energy, the possible deviations with respect to special relativity could be studied in the near future. In the present work we have considered, for the first time, possible consequences from a perspective of a deformed kinematics that is relativistic invariant, in contrast to the alternative, already explored~\cite{Li:2021cdz,Li:2021tcw,Chen:2021hen,Satunin:2021vfx,LHAASO:2021opi}, LIV case.

The conclusion of the studies of possible LIV effects in LHAASO observations is that one has a sensitivity to a LIV scale of the order of, or above, the Planck scale. In contrast, in the example of a relativistic deformed kinematics considered in this work, we have shown that one has a sensitivity to a scale thirteen orders of magnitude below the Planck scale. The dependence on the energy of the modification of the mean free path in the case of LIV differs from the case of a relativistic deformed kinematics. In fact, while a subluminal LIV produces necessarily a greater transparency of the Universe than in SR, the curves in Fig.~\ref{fig:figura8} are above or below the SR mean free path for different energy ranges. Then, in case future data would allow us to identify a modification in the transparency of the Universe expected in SR, the spectrum of very high-energy gamma rays could allow to distinguish if the origin of the modification is a LIV or a deformed relativistic kinematics.

Finally, let us note that the model of relativistic deformed kinematics considered in this work allowed us to carry out the computation of the mean free path  in a very simple way from the result in the SR case. It would be interesting to consider the mean free path with other models of relativistic deformed kinematics. The comparison of results could clarify the relation between different models of relativistic deformed kinematics, which is still an open question in the DSR  scenario~\cite{KowalskiGlikman:2002we,Carmona:2017oit}.  

We limited ourselves in this work to the local Universe and to an energy range where the effect of the electromagnetic background on the propagation of high-energy photons is dominated by the CMB. This is relevant for the highest energies which are being and will be explored by LHAASO and next generation expriments, around and a few orders of magnitude above the PeV scale. However, following the same strategy of this paper, one could extend the energy range considered taking into account the photon density corresponding to each  background of low-energy photons. Moreover, the study of the effects of a relativistic deformed kinematics presented in this work can be generalized to other processes, opening the window to explore new astroparticle effects in the DSR framework.

\section*{Acknowledgments}
This work is supported by Spanish grants PGC2018-095328-B-I00 (FEDER/AEI), and DGIID-DGA No. 2015-E24/2. JJR acknowledges support from the INFN Iniziativa Specifica GeoSymQFT and
Unión Europea-NextGenerationEU (``Ayudas Margarita Salas para la formación de jóvenes doctores'').  The work of MAR was supported by MICIU/AEI/FSE (FPI grant PRE2019-089024). This work has been partially supported by Agencia Estatal de Investigaci\'on (Spain)  under grant  PID2019-106802GB-I00/AEI/10.13039/501100011033. The authors would like to acknowledge the contribution of the COST Action CA18108 ``Quantum gravity phenomenology in the multi-messenger approach''.

\end{document}